\documentclass[pra,twocolumn,showpacs,amssymb,aps]{revtex4}
\usepackage{dcolumn}
\usepackage{bm}
\usepackage{graphicx}
\usepackage{epstopdf}
\usepackage{subfig}
\usepackage{color}
\usepackage{amsmath}
\usepackage{tikz-cd}
\usetikzlibrary{decorations.pathreplacing}
\usepackage[colorlinks=true]{hyperref}
\newcommand*\sech{\mathop{}\!\operatorname{sech}}

\newcommand*\cn{\mathop{}\!\operatorname{cn}}
\newcommand*\dn{\mathop{}\!\operatorname{dn}}
\newcommand*\am{\mathop{}\!\operatorname{am}}

\begin{document}
	\title{The Complex Korteweg-de Vries Equation: A Deeper Theory of Shallow Water Waves}
	\author{M. Crabb and N. Akhmediev}
	\affiliation{Department of Theoretical Physics, Research School of Physics, The Australian National University, Canberra, ACT, 2600, Australia}
	
	\begin{abstract} 
	Using Levi-Civit\`a's theory of ideal fluids, we derive the complex Korteweg-de Vries (KdV) equation, describing the complex velocity of a shallow fluid up to first order. We use perturbation theory, and the long wave, slowly varying velocity approximations for shallow water. The complex KdV equation describes the nontrivial dynamics of all water particles from the surface to the bottom of the water layer. A crucial new step made in our work is the proof that a natural consequence of the complex KdV theory is that the wave elevation is described by the real KdV equation. The complex KdV approach in the theory of shallow fluids is thus more fundamental than the one based on the real KdV equation. We demonstrate how it allows direct calculation of the particle trajectories at any point of the fluid, and that these results agree well with numerical simulations of other authors.
	\end{abstract}
	\pacs{05.45.Yv, 42.65.Tg, 42.81.qb 47.35.-i}
	\maketitle
	\date{today}
	\section{Introduction}
Water waves can be classified into many types \cite{Toffoli}. One shared feature is, as Feynman put it, that they have all the possible complications that a wave can have. For instance, when treated with full generality, water waves are commonly considered to be nonlinear phenomena \cite{Constantin}. Consequently, to precisely describe water waves is infamously difficult.  One convenient approximate method of describing water waves is to give an evolution equation for the elevation of the water surface \cite{Sedlet,Slunyaev}; in the lowest order nonlinear approximation, this leads to the nonlinear Schr\"odinger equation for the complex envelope of waves in deep water \cite{Osborne}, and the real Korteweg-de Vries (KdV) equation for the elevation of a nonlinear shallow water wave \cite{Boussinesq, Korteweg, Grimshaw}. Higher-order physical effects can be incorporated in each of these equations for more descriptive power \cite{Slunyaev,Dysthe,Trulsen,Marchant}. However, these approaches still restrict us to only the motion of the surface.\\\indent
When a fluid is incompressible, we can also describe its state of motion with the potential. As is often detailed in many standard texts on hydrodynamics \cite{Milne-Thomson}, a well-behaved potential can be treated as the independent variable in the description of the fluid, rather than a spatial coordinate. This approach was pioneered by Levi-Civit\`a in 1907, who derived an equation satisfied by all fluid motions without singularities in a channel of arbitrary depth \cite{Levi-Civita}. The theory was developed in the early twentieth century, but due to the difficult nature of the resulting equation it has fallen by the wayside. Modern interest in this approach has been partly revived by Levi \cite{Levi}.\\\indent
In this work, we give a thorough derivation of a complex KdV equation describing first order perturbations in the complex velocity around steady flow. Importantly, not only the dependent function in the KdV equation but the spatial variable is considered to be complex as well. This approach provides a complete description of the flow in the fluid up to first order, and is not limited to describing only the water elevation. Thus, the main advantage of the complex
KdV equation in hydrodynamics is that it describes the dynamics of water particles not only at the surface but also throughout the entire body of the fluid.\\\indent
A new step forward in our work is that if the complex KdV equation describes a first-order perturbation of the complex velocity for a shallow fluid, then the fluid's elevation is naturally determined by a solution of the standard real KdV equation.\\\indent
As an example application of this theory, we give a simple demonstration of how the motion of the entire fluid may be described with a basic periodic solution to the complex KdV equation. We illustrate how the particle displacement from a given point may be determined, along with their trajectories, and show that in the limit as the period becomes infinite, the familiar soliton solution may be correctly described. 

\section{The Levi-Civit\`a Theory of Potential Flow in Inviscid Fluids}
\begin{figure}[ht]
\begin{tikzpicture}
\draw[->]   (0,-1) -- (7.5,-1);
\draw[thick, gray] plot[domain=0:7.3*pi, samples=300]  (\x/pi,{1+sin(\x r/3)}); 
\draw[densely dashed]   (0,1) -- (7.5,1);
\draw[densely dashed]   (0,2.03) -- (7.5,2.03);
\draw[->] (3.75,-1) coordinate (O) -- (3.75,2.5);
\node [below] at (7.2,-1) {$x$};
\node [right] at (3.75,2.2) {$y$};
\coordinate (h) at (5.86,0.98);
\coordinate (h') at (6.1,1.98);
\draw[decoration={brace,mirror,raise=5pt},decorate]
(5.86,-0.97) -- node[right=6pt] {$h$} (h);
\draw[decoration={brace,raise=5pt},decorate]
(6.1,1.02) -- node[left=6pt] {$\eta(x,t)$} (h');
\end{tikzpicture}\\ 
\caption{\it Schematic of water layer with average depth h. The function $\eta(x,t)$ describes the water elevation above the average level. \label{illustration}}
\end{figure}
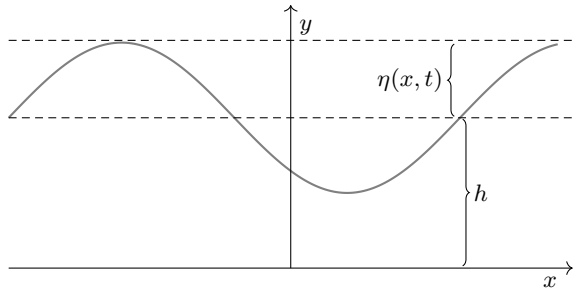

Water of depth $h$ occupies a channel with a flat bottom, and waves of a height $\eta=\eta(x,t)$ above the mean level propagate along the surface, the axis of $x$ being taken along the bottom, and in the direction of propagation, while $y$ represents a vertical coordinate. A diagram is shown in Fig.\ref{illustration}.
At any point $x$ at a time $t$ the water surface is described by the equation $y=h+\eta(x,t).$ Thus,
\[\frac{\partial\eta}{\partial t}+u\frac{\partial\eta}{\partial x}=v,\]
where $u$ and $v$ are the horizontal and vertical components of the fluid velocity, respectively.\\\indent
When the ratio of wave height to wavelength is very small, we have the linearised theory
\[\frac{\partial\eta}{\partial t}=\frac{\partial\psi}{\partial x}\]
where $\psi$ is the stream function, with the convention $d\psi=vdx-udy$. If we additionally suppose that the motion is irrotational, and that the wave is long enough that surface tension can be neglected, then the pressure just inside the fluid's surface must be very nearly equal to the pressure just outside the fluid's surface, and this implies that, approximately,
\[g\eta\simeq\frac{\partial\phi}{\partial t},\]
where $\phi$ is the velocity potential, with $d\phi=-udx-vdy$.\\\indent
Since there is no flow through the bottom of the channel, the stream function is constant on the bottom of the channel, and we can choose $\psi=0$ when $y=0$. The complex potential $w=\phi+i\psi$ is then real when $y=0,$ and $w$ can be analytically continued into the region $-h\leqslant y<0.$ Doing so leads to Cisotti's equation,
\begin{align}
\label{cisotti}
&\frac{\partial^2}{\partial t^2}\left\{w(z+ih,t)+w(z-ih,t)\right\}+\nonumber\\
&~~~+ig\frac{\partial}{\partial z}\left\{w(z+ih,t)-w(z-ih,t)\right\}=0,\end{align}
More details can be found in the well-known book by Milne-Thomson \cite{Milne-Thomson}. This equation is complex but linear.
However, we are also interested in nonlinear waves.\\\indent
Beginning from just Bernoulli's principle, if we let $q=\sqrt{u^2+v^2}$ be the total speed of the fluid at a given position and time, then
\[-\frac{\partial\phi}{\partial t}+\tfrac12q^2+gy+\frac{p}{\rho}=f(t)\]
where $\rho$ is the density of the fluid, $p$ is the pressure, and the time-derivative of the velocity potential is the energy due to acceleration within the fluid.\\\indent
Now we suppose that the fluid surface is free to move along a variable curve given by $y=h+\eta(x,t).$ Along the free surface, the pressure is constant, so we have
\begin{equation}
\label{bernoulli}-\frac{\partial\phi}{\partial t}+\tfrac12q^2+g\eta=0.
\end{equation}
We can define the velocity potential $\phi$ such that any constant or function of time on the right hand side can be set to zero. Note also that for surface waves, $0\leqslant|\psi|\leqslant|\psi_0|,$ where $\psi=\psi_0$ on the free surface $y=h+\eta(x,t).$\\\indent
Let us denote $w=\phi+i\psi,$ $z=x+iy,$ and $\Upsilon=u-iv$. We reiterate
\begin{equation}
\label{cv}
\Upsilon=-\frac{\partial w}{\partial z}.
\end{equation}
Since $w$ and $\Upsilon$ are both real along the bottom of the channel, and certainly assumed to be holomorphic throughout the body of fluid, we can analytically continue both functions to the region where $-h\leqslant y<0$. Also, since $w$ is holomorphic at every point of the flow, we can invert the relationship to write $z,$ and the complex velocity $\Upsilon,$ as functions of the potential $w$ rather than position. The velocity potential and stream function satisfy, by (\ref{cv}), the differential equation
\[dx+idy=-\frac{d\phi+id\psi}{u-iv}.\]
Streamlines are defined by $d\psi=0,$ so, with 
\[ds^2=dx^2+dy^2\] 
defining the line element on the free surface, and after collecting real and imaginary parts, which are respectively
\begin{align*}
dx&=-\frac{ud\phi-vd\psi}{q^2},\\
dy&=-\frac{vd\phi+ud\psi}{q^2},
\end{align*}
we have
\[\frac{\partial y}{\partial s}=-\frac{v}{q^2}\frac{\partial\phi}{\partial s}\]
on a streamline. We also have, by definition, 
\[-d\phi=udx+vdy,\]
so it is easy to see that
\[\frac{\partial\phi}{\partial s}=-q,\]
and consequently
\[\frac{\partial\eta}{\partial s}=\frac{v}{q}\]
on a streamline. Bernoulli's equation (\ref{bernoulli}) can now be differentiated with respect to an arc length $s$ along the free surface $y=h+\eta(x,t)$ to give
\[\frac{\partial q}{\partial t}+q\frac{\partial q}{\partial s}+\frac{gv}{q}=0,\]
or
\begin{equation}
\label{b2}
\frac{\partial q^2}{\partial t}+q\frac{\partial q^2}{\partial s}+2gv=0.\end{equation}
As stated only shortly before, we can recast Bernoulli's principle, now in the form (\ref{b2}), in terms of the complex velocity $\Upsilon$ and the complex potential $w.$ We obtain through a simple change of variable $d\phi=-qds$ the equation
\begin{equation}
\label{b3}
\frac{\partial}{\partial t}|\Upsilon|^2-|\Upsilon|^2\frac{\partial}{\partial\phi}|\Upsilon|^2+ig(\overline{\Upsilon}-\Upsilon)=0.
\end{equation}
Here, instead of the condition that the free surface be defined by $y=h+\eta(x,t)$, we have instead the free surface defined by the streamline $\psi=\psi_0,$ and the complex velocity should be understood as a function $\Upsilon=\Upsilon(\phi+i\psi_0,t),$ with the conjugate velocity $\overline{\Upsilon}$ given by $\overline{\Upsilon}=\Upsilon(\phi-i\psi_0,t).$ Lastly, since $\psi=0$ and therefore $w$ is real along the bottom of the channel, $y=0,$ we can extend the differential equation for $\Upsilon$ to all $w$, rather than just for $w=\phi.$ The differential equation (\ref{b3}) can therefore be presented in the form
	\begin{eqnarray}
\nonumber
	&-&\frac{\partial}{\partial t}\log\{\Upsilon(w+i\psi_0,t)\Upsilon(w-i\psi_0,t)\} +
	\\ \nonumber
	&+&\frac{\partial}{\partial w}\{\Upsilon(w+i\psi_0,t)\Upsilon(w-i\psi_0,t)\}+ \\
	&+& ig\left\{\frac{1}{\Upsilon(w+i\psi_0,t)}-\frac{1}{\Upsilon(w-i\psi_0,t)}\right\} = 0.
	\label{lce}	\end{eqnarray}
This equation was first derived by Levi-Civit\`a \cite{Levi-Civita} over a century ago, albeit in a more limited form, dealing only with steady flow. However, due to the fact that a differential equation of this type is difficult to solve, there has been only relatively minor attention given to this particular equation. Having said that, even though this equation is challenging to solve in the general case, it is nonetheless possible to gain insight into possible fluid dynamics by applying certain techniques, such as perturbation series.
\section{Linear Perturbations around a Steady Flow}
Suppose that the complex velocity of the wave has the form of a small perturbation around an otherwise constant flow parallel to the bottom of the channel, so
\[\Upsilon(w,t)=c\{1+\varepsilon\alpha(w,t)\},\]
where $\varepsilon$ is a small parameter, $\alpha$ is a complex function of the potential and $c$ is a real constant. Since we are restricting ourselves to considering only small perturbations around a steady horizontal flow, most of the flow across the surface at any given point will be due to the horizontal movement, and therefore we can take the flux over the free surface to be approximately $\psi_0=-ch.$ 
This will clearly be true for periodic functions, but also for the general problem of the solitary wave \cite{Milne-Thomson}, since the fluid must return to steady motion infinitely far away from the travelling pulse.\\\indent
By hypothesis, $\varepsilon^2\simeq0,$ so we can disregard those terms of all but first order. To first order, the perturbation satisfies
\begin{align}
\label{perturbation}
&-\frac{\partial}{\partial t}\log[1+\varepsilon\{\alpha(w+ich,t)+\alpha(w-ich,t)\}]+\nonumber\\
&~~~+c\varepsilon\frac{\partial}{\partial w}\left\{\alpha(w+ich,t)+\alpha(w-ich,t)\right\}+\nonumber\\
&~~~+\frac{ig\varepsilon}{c^2}\left\{\alpha(w+ich,t)-\alpha(w-ich,t)\right\}=0.
\end{align}
This is the equation which will determine the stability of a small perturbation in the velocity.\\\indent
For simplicity, first, we will consider motion which can be reduced to steady flow, such as, for example, a solitary wave or periodic motion. When the motion is steady, all the time dependence will be implicit, and contained in the potential, so that (\ref{perturbation}) becomes
\begin{equation}
\label{po}
\left\{\frac{d}{dw}\cos\left(ch\frac{d}{dw}\right)-\frac{g}{c^3}\sin\left(ch\frac{d}{dw}\right)\right\}\alpha(w)=0.
\end{equation}
Here, the sine and cosine terms should be understood, like the exponential, in the sense of formal power series of an operator. That is,
\begin{equation}\cos\left(\frac{d}{dw}\right)=\sum_{n=0}^{\infty}\frac{(-1)^n}{(2n)!}\frac{d^{2n}}{dw^{2n}},\end{equation}
and similar for the sine operator.\\\indent
The ease of manipulation which results from the form (\ref{po}) immediately suggests a simple solution for steady flow in the form
\begin{equation}
\label{steadyperturbation}
\alpha(w)=\alpha_0e^{i \mu w},
\end{equation}
where $\mu$ is a solution of the transcendental equation
\begin{equation}
\label{transcendental}
\tanh ch\mu=\frac{c^3\mu}{g},
\end{equation}
and $\alpha_0$ a constant. If $\alpha_0$ is small, integration gives the equation
\begin{equation*}
	z=-\frac1c\left\{w+\frac{i\alpha_0}{\mu}e^{i\mu w}+O(\alpha_0^{\;2})\right\}
\end{equation*}
and thus
\begin{equation}
	\eta=\frac{\alpha_0}{\mu}e^{ch\mu}\cos\mu cx+O(\alpha_0^{\;2})
\end{equation}
for the stationary form of the free surface, up to first order. We see that $\mu=k/c,$ approximately, where $k$ is the wavenumber of the flow, and that (\ref{transcendental}) is just the dispersion relation for waves in an inviscid fluid of depth $h.$\\\indent
Now suppose the motion is steady. We can write the potential purely as a function of $z$; $w=w(z),$ and to lowest order in $\varepsilon$, we have 
\[w=-cz+O(\varepsilon).\]
To lowest order in the $z$-plane:
\begin{equation}
\label{o0c}
-\frac{d}{dz}\{\beta(z+ih)+\beta(z-ih)\}=\frac{ig}{c^2}\{\beta(z+ih)-\beta(z-ih)\}.	
\end{equation}
This is effectively equivalent to Cisotti's equation (\ref{cisotti}).\\\indent
For the case in which the channel is shallow, with a flat bottom, we can obtain an approximate equation of motion by expanding in powers of $h$, which will be a convergent series when $h$ is sufficiently small. 
\section{The Nonlinear Theory}
We can more generally consider a perturbation series in powers of the small parameter $\varepsilon$ of the form
\begin{equation}
\label{series}
\Upsilon=c(1+\varepsilon\beta_1+\varepsilon^2\beta_2+\cdots+\varepsilon^n\beta_n+\cdots),
\end{equation}
where $\beta_n=\beta_n(z,t).$ We can expand (\ref{lce}) to each order of $\varepsilon,$ and solve the equations obtained at each order sequentially. The constant flow velocity $c$ will have distinct values for the shallow and deep regimes, and these must be treated as two separate limiting cases to make the problem manageable. 
\subsection{The Complex KdV Equation}
For shallow fluids, we consider perturbations around a constant flow $c=\sqrt{gh}.$ We can expand $\Upsilon(w\pm ich,t)$ in powers of $h,$ since $h$ is small. We then apply the change of variable $\Upsilon(w,t)\mapsto\Upsilon(z,t)$, under which the derivative transforms as
\begin{equation}
\label{change}
\frac{\partial\Upsilon}{\partial w}\mapsto-\frac{\partial}{\partial z}\log\Upsilon(z,t).
\end{equation}
Higher order derivatives transform as
\begin{align*}
\frac{\partial^2\Upsilon}{\partial w^2}&\mapsto\frac{1}{\Upsilon}\frac{\partial^2}{\partial z^2}\log\Upsilon,\\
\frac{\partial^3\Upsilon}{\partial w^3}&\mapsto-\frac{1}{\Upsilon^2}\frac{\partial^3}{\partial z^3}\log\Upsilon+\frac{1}{\Upsilon^3}\frac{\partial\Upsilon}{\partial z}\frac{\partial^2}{\partial z^2}\log\Upsilon,\end{align*} and so on. Truncating the series at third order gives the approximate equation
\begin{align*}-\Upsilon^2\frac{\partial\Upsilon}{\partial z}+(gh)^{\frac32}\frac{\partial}{\partial z}\log\Upsilon + &
\\ 
+\frac{gh^3}{2}\bigg(\frac{\partial^3\Upsilon}{\partial z^3}-3\frac{1}{\Upsilon}\frac{\partial\Upsilon}{\partial z}\frac{\partial^2\Upsilon}{\partial z^2}\bigg)+ &
\\
+\frac{g^{\frac52}h^{\frac92}}{6}\left(\frac{1}{\Upsilon^3}\frac{\partial\Upsilon}{\partial z}\frac{\partial^2}{\partial z^2}\log\Upsilon-\frac{1}{\Upsilon^2}\frac{\partial^3}{\partial z^3}\log\Upsilon\right)& \simeq2\Upsilon\frac{\partial\Upsilon}{\partial t}\end{align*}
after a change of variable to the $z$-plane. We then assume a perturbation series solution of the form
\[\frac{\Upsilon(z,t)}{\sqrt{gh}}=\sum_{n=0}^{\infty}\varepsilon^n\beta_n(z,t).\]
The lowest order term is identically zero with this substitution, as we have seen. The only surviving coefficient of $\varepsilon$ gives the first order approximation
\[\frac{2(gh)^{\frac32}}{3}\left(h^2\frac{\partial^3\beta_1}{\partial z^3}-\frac{3}{\sqrt{gh}}\frac{\partial\beta_1}{\partial t}\right)\varepsilon,\]
while the second-order terms read
\begin{align*}\bigg\{- 6\beta_1\frac{\partial\beta_1}{\partial z}-\frac{5h^2}{3}\frac{\partial\beta_1}{\partial z}\frac{\partial^2\beta_1}{\partial z^2}+h^2\beta_1\frac{\partial^3\beta_1}{\partial z^3}-&
\\ 
-\frac{2}{\sqrt{gh}}\beta_1\frac{\partial\beta_1}{\partial t}
+\frac{2h^2}{3}\frac{\partial^3\beta_2}{\partial z^3}-\frac{6}{\sqrt{gh}}\frac{\partial\beta_2}{\partial t} & \bigg\}(gh)^{\frac32}\varepsilon^2.\end{align*}
Since the same stray terms that appear as the coefficients of $\varepsilon$ appear as coefficients of $\varepsilon^2$ with $\beta_2$ instead of $\beta_1,$ it is reasonable to suspect that these terms actually belong to a higher order of smallness.\\\indent
If we let a typical wave in the fluid have a characteristic length scale $l,$ then to say that the fluid is shallow is only the statement that the fraction $h/l$ is much less than 1. For a wave which is typically long, the wave will normally appear to change only slowly, and over a long horizontal distance of the same order as $l.$ If we put
\[\left(\frac{h}{l}\right)^2=\delta,\]
where $\delta$ is small, then this suggests defining the dimensionless variable 
\[Z=X+iY=\frac{z}{l}.\]
The condition that the fluid is shallow is now that the vertical coordinate $Y$ is much smaller than 1 everywhere in the fluid; more specifically, $Y=O(\delta^{\frac12})$. We will also assume that $\delta=\varepsilon.$ \\\indent
Writing equations in terms of $Z,$ we see that the variation of the first order perturbation $\beta_1$ must be infinitesimal with respect to $t.$ This is physically intuitive since the characteristic length scale of the wave is large. However, upon the introduction of a longer time scale, $t'=\varepsilon t,$ $\beta_1$ can be seen to have a nontrivial slow variation; we have at first order a complex Korteweg-de Vries (KdV) equation:
\[\frac{l}{\sqrt{gh}}\frac{\partial\beta_1}{\partial t'}=-3\beta_1\frac{\partial\beta_1}{\partial Z}+\frac{1}{3}\frac{\partial^3\beta_1}{\partial Z^3}.\]
If we take the ratio of $t'$ to the characteristic time $l/\sqrt{gh}$, we can define a dimensionless long time scale $T$ by
\[\frac{t'}{T}=\frac{l}{\sqrt{gh}}.\]
In terms of $Z$ and $T,$ we have the fully dimensionless form of the complex KdV equation,
\begin{equation}
\label{ckdv}
\frac{\partial\beta_1}{\partial T}=-3\beta_1\frac{\partial\beta_1}{\partial Z}+\frac{1}{3}\frac{\partial^3\beta_1}{\partial Z^3}.
\end{equation}
The conditions of validity are that the wave is generally long, and the perturbation varies slowly with time.\\\indent
The idea of reviving the theory of the complex KdV equation comes from the work of Levi \cite{Levi}. However, some fundamental errors made in previous work \cite{Levi, Levi2} did not allow this theory to be used in practice. Namely, in \cite{Levi}, the water elevation $h$ was directly replaced by the expression $h=h_0\varepsilon^{\frac13}$ with $h_0$ being an unspecified parameter. The smallness parameter introduced this way is incompatible with approximations required for establishing correspondence with the real KdV equation. In the work \cite{Levi2}, an intended small parameter  $\varepsilon$ was not dimensionless, making the idea of smallness in a perturbation series poorly defined. Here, by instead scaling the complex variable by a characteristic length scale $l$, the typical length of the wave enters through the derivative terms, providing a physically clear interpretation of how small each of the terms are. The two scales of the flow, $h$ and $l$, then naturally form the dimensionless parameter $\varepsilon=(h/l)^2,$ which is also the same order of smallness as in the real KdV equation \cite{Toda}, up to a possible constant factor.\\\indent
Unlike the better-known real KdV equation \cite{Korteweg}, the complex KdV equation describes a perturbation of the complex velocity around an otherwise steady, uniform flow. It thus describes the entire fluid motion, not just the elevation of the free surface. Consequently, our approach allows us to describe the trajectories of the fluid particles across the whole layer of liquid. This has not actually been carried out at all up until now, although this is one of the important advantages of the complex KdV equation. Below, we provide the first demonstration of how this is done, as well as point out its excellent qualitative agreement with numerical results of other authors.\\\indent
But before we do, we will show, for the first time, how the complex KdV theory leads naturally to the standard theory of the real KdV equation for the surface elevation of shallow water.
\subsection{Relation to the Real KdV Equation}
We can construct a similarly dimensionless potential $\tilde{w}$ by the scaling
\begin{equation}
\label{dimensionlessshallowpotential}
\tilde{w}=\frac{w}{\sqrt{gh}l}=-Z+\varepsilon\gamma_1+\varepsilon^2\gamma_2+\cdots
\end{equation}
The corresponding first order perturbation of the potential in dimensionless form will give $\beta_1$ through
\begin{equation}
\label{potentialperturbation1}
-\frac{\partial\gamma_1}{\partial Z}=\beta_1(Z,T).
\end{equation}
Substituting this into (\ref{ckdv}) gives the potential KdV equation for the function $\gamma_1$:
\begin{equation}
\label{pkdv}
	\frac{\partial\gamma_1}{\partial T}=\frac32\left(\frac{\partial\gamma_1}{\partial Z}\right)^2+\frac{1}{3}\frac{\partial^3\gamma_1}{\partial Z^3}.
\end{equation}

Equation (\ref{pkdv}) also gives one of the simplest methods for obtaining the elevation $\eta$. We have assumed that, whatever the form of the free surface, it corresponds to the constant value of the stream function $\psi=-ch.$ The imaginary part of the dimensionless potential $\tilde{w}=\tilde{\phi}+i\tilde{\psi}$ on the surface is then just
\[\tilde{\psi}=-\frac{h}{l}.\]
Suppose that a holomorphic solution for the first order perturbation of the potential $\gamma_1=\gamma_1(Z,T)$ is obtained. If $\gamma_1$ has imaginary part $\tilde{\psi}_1,$ then we must have on the free surface $y=h+\eta$ the equation
\begin{equation}
\label{sfsurface}
\frac{\eta}{l}\simeq\varepsilon\tilde{\psi}_1|_{Y=(h+\eta)/l}
\end{equation}
in terms of the dimensionless variables, accurate to first order in $\varepsilon$. In general, this will give an implicit equation for $\eta$.\\\indent
In fact, since $Y$ has already been required to be small, of order $\sqrt{\varepsilon},$ we can avoid the complications of implicit equations for the free surface by approximating the first order perturbation of the potential as
\[\gamma_1(Z,T)\simeq\gamma_1(X,T)-iY\beta_1(X,T).\]
Recalling that the stream function vanishes on the bottom of the channel, it follows that $\gamma_1(X,T)$ is real, so the first order perturbation in the stream function must be approximately
\[\tilde{\psi}_1\simeq-Y\beta_1(X,T).\]
From (\ref{sfsurface}), the surface elevation $\eta$ can be given in terms of $\beta_1(X,T)$ as
\[\eta(X,T)\simeq-\varepsilon\{h+\eta(X,T)\}\beta_1(X,T),\]
or
\begin{equation}
\label{elevationkdv}
\eta(X,T)\simeq-h\varepsilon\beta_1(X,T)
\end{equation}
to first order in $\varepsilon$.\\\indent
The vanishing of the stream function along the bottom is equivalent to the vanishing of the vertical component of the velocity. We see then that $\beta_1(X,T)$ must be real.\\\indent
Given that $\beta_1(X,T)$ is a real solution to (\ref{ckdv}) with the real variable $X$ instead of the complex variable $Z,$ we have obtained in (\ref{elevationkdv}) the result that the surface elevation $\eta$ of a shallow fluid is described by a real solution of the KdV equation in $X$ and $T$, up to a constant of proportionality, and accurate to first order in $\varepsilon$.\\\indent
We therefore make the claim that the complex KdV equation is more fundamental than the real KdV equation, since we have shown that the description of the surface elevation by the real KdV equation naturally emerges as a consequence of a perturbation of the complex velocity being described by the complex KdV equation. The complex KdV equation, however, also allows for a full picture of the motion of the entire fluid up to first order, not just the surface elevation to which we are limited by considering only its real solutions. Not only mathematically, then, but also physically, we have thus justified the claim that it is the complex KdV equation which takes the most fundamental place in the hydrodynamic theory of shallow fluids.\\\indent
It is also worth mentioning that the $\tau$-function for the complex KdV equation, which provides maybe the simplest mathematical lens through which to view the KdV equation \cite{Hirota}, can be simply related to the first order perturbation of the potential, since
\begin{equation}
	\log\tau(Z,T)\propto\int^Z\gamma_1(Z',T)dZ'.
\end{equation}
In the complex KdV equation, the $\tau$-function is thus a logarithmic measure of flux in the fluid due to perturbations around steady flow. To develop any more sophisticated physical connection between the complex potential and the $\tau$-function further would be beyond the scope of this work, so we leave this as what we believe to be an interesting comment.\\\indent
Terms of the third order of smallness give an equation for $\beta_2.$ Naturally, the calculations are more complicated for higher orders of the perturbation series, and are less immediately enlightening, so we relegate these to the appendix. However, we stress that higher-order corrections to the wave motion can be naturally calculated in an extended version of this theory.

\section{Particle Motion in Periodic Waves}
\begin{figure}[ht]\includegraphics[scale=0.4]{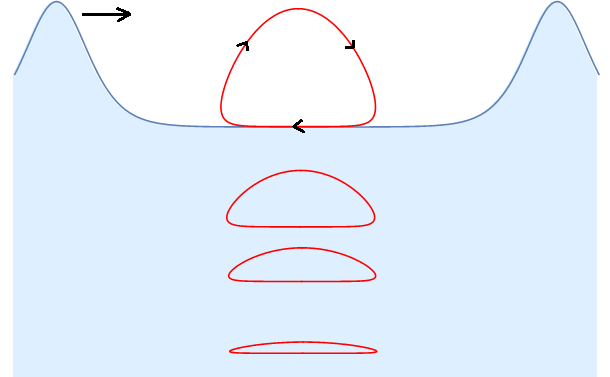}
	\caption{\it An illustration of the paths of the particles in a rightward-propagating cnoidal wave (not to scale). The surface elevation is described by the real KdV equation, but the motion of the particles within the fluid is described by the complex KdV equation (\ref{ckdv}). The motion of the particle at a given point is determined by (\ref{dispv}, \ref{disph2}). The trajectory depends on the wave modulus, frequency, and depth in the fluid.}
\label{sketch2}
\end{figure}

Ignoring the constant background flow, from a fixed point $Z$ a fluid particle will move to the point $Z+dZ'$ in a time $dT$ under the influence of the perturbation $\beta_1.$ In this section, we will drop the subscript without confusion, since we are only working to first order in $\varepsilon,$ in the KdV regime.\\\indent
The rate of change of the particle's position will be given by the differential equation
\begin{equation}
\frac{d\overline{Z}'}{dT}=\beta,
\end{equation}
where, again, the bar denotes complex conjugation.\\\indent
A periodic wave will have the form $\beta=\beta(mZ-nT).$ 
Integrating with respect to $T$ gives
\begin{equation}
\overline{Z}'=\frac{\gamma(mZ-nT)}{n}.
\end{equation}
As before, we can make the lowest order approximation
\[\gamma(mZ-nT)\simeq\gamma(mX-nT)-imY\beta(mX-nT),\]
so that the new position of the particle is given by the horizontal and vertical coordinates
\begin{align*}
	X'&=\frac1n\gamma(mX-nT)+C_1,\\
	Y'&=\frac{m}{n}Y\{\beta(mX-nT)+C_2\},
\end{align*}
where $C_1$ and $C_2$ are constants of integration (when seeing that a term of the form $C_2Y$ can appear, it must be remembered that the initial position $(X,Y)$ is fixed).\\\indent
To be definite, we will consider a particular periodic solution of the complex KdV equation. From (\ref{ckdv}), we see that we have a periodic solution of the form
\begin{equation}\label{cn}
\beta(Z,T)=-\tfrac43k^2m^2\cn^2(mZ+nT,k),
\end{equation}
where $k$ is the modulus while the frequency is
\[n=-\tfrac43(2k^2-1)m^3.\]
We also have one of the form
\begin{equation}
\beta(Z,T)=-\tfrac43m^2\dn^2(mZ+nT,k),
\end{equation}
where
\[n=-\tfrac43(k^2-2)m^3,\]
and another of the form $\cn^2+k\cn\dn$ \cite{KhareSaxena}. As $k\to1,$ the first two tend to the same $\sech^2$-type solution, while the third tends to 0. Also, because of the identity
\begin{equation}
\label{dncn}
\dn^2(u,k)=k'^2+k^2\cn^2(u,k),
\end{equation}
where $k'$ is the complementary modulus, the $\cn^2$ and $\dn^2$-type solutions can be simply transformed into one another through Galilean symmetry. We will therefore use the term cnoidal wave to refer to both of these solutions.\\\indent
Corresponding to the $\dn^2$-solution, the perturbation in the potential is
\begin{equation}
\gamma(Z,T)=-\tfrac43mE(\am\zeta,k),
\end{equation}
where $E(u,k)$ is the elliptic integral of the second kind, and $\zeta=mZ-nT.$ If we look at the $\cn^2$-solution instead, $\gamma$ just picks up an extra term which is just a multiple of $\zeta$ from the relation (\ref{dncn}). Again, because $Z$ refers to a constant position in the fluid, not a coordinate which follows the particle, this will not lead to qualitatively different behaviour up to Galilean symmetry.\\\indent
The coordinates of the fluid particle are given by
\begin{align}
\label{disph1}
	X'&=-\frac{4m}{3n}E(\am\xi,k)+C_1,\\
\label{dispv}
	Y'&=-\frac{4m^3Y}{3n}\{\dn^2(\xi,k)+C_2\},
\end{align}
where $\xi=mX-nT.$ By use of the identity
\[E(\am u,k)=Z(u,k)+\frac{Eu}{K},\]
where $K=K(k)$, $E=E(k)$ are the complete elliptic integrals of the first and second kind, respectively, and $Z(u,k)$ is the Jacobi zeta function with modulus $k,$ we can write
\begin{equation}
\label{disph2}
	\xi'-\xi'_0=-\frac{4m^2}{3n}Z(\xi,k).
\end{equation}
Here $\xi'$ represents horizontal position in the comoving frame and $\xi'_0$ is a fixed constant. The fluid motion is now easily seen to be periodic, with identical motion at points separated by a horizontal distance $2K$ in the comoving frame, i.e. $\xi\sim\xi+2K.$\\\indent
The trajectory of any particle in the fluid must be a closed curve (up to the constant horizontal flow). However, it cannot be the usual ellipse, which is characteristic of linear flow. It is easy to see this by physical principles alone. Because cnoidal waves are characterised by long, relatively flat regions between narrow peaks, the trajectory of a particle cannot be symmetric in two axes. Instead, it will travel along a nearly flat line, and then its vertical velocity will increase and decrease quickly. This means that its trajectory must be a curve which has a broad, flat base. The width of this curve will also increase with the modulus $k,$ and be constant with respect to $Y,$ since $\xi'$ is independent of $Y$. The height of the trajectory will also decrease until it is completely flat at the bottom of the channel.\\\indent
We give an illustration of the periodic motion of a cnoidal wave in the complex KdV equation in Fig.\ref{sketch2}, taking $k$ large enough that the flat regions between peaks becomes clearly defined. The surface has the form of a cnoidal wave, while the curves on and within the fluid display the trajectories of fluid particles as we descend to the bottom. 
The trajectories in the cnoidal wave are most similar to bean curves, rather than ellipses or circles, which are familiar from the linear theory. However, when $k$ is small enough, the trajectories may be well-approximated by ellipses. The particle trajectories are in close agreement with earlier numerical approximations \cite{Borluk}, but in the framework of the complex KdV theory, they appear naturally with very little work.\\\indent
As $k\to1,$ we have also $K\to\infty,$ so the motion is no longer periodic. Instead, we have the well-known $\sech^2$-shaped soliton solution. In this case, the particle trajectories become parabolic arcs \cite{McCowan}.\\\indent
Lastly, we will give an example calculation of the elevation from the complex KdV equation. The simplest case to work with is that of a first order perturbation in the potential of the form
\begin{equation}
\label{tanh}
\gamma=\tfrac43a\tanh a(Z+\tfrac43a^2T).
\end{equation}
This leads to the implicit equation for the surface
\[\frac{\eta}{l}\simeq\frac{\frac43\varepsilon a\sin2a(h+\eta)/l}{\cos2a(h+\eta)/l+\cosh2a(X+\tfrac43a^2T)}.\]
An initial measurement taken at $T=0$ will have a maximum of $\eta_0$ located at $X=0,$ roughly given by the transcendental equation
\[\frac{\eta_0}{l}\simeq\tfrac43\varepsilon a\tan\left(a\frac{h+\eta_0}{l}\right).\]
When $h$ and $l$ are known and the dimensionless parameter $a$ is fixed, this allows an approximation of $\eta_0.$ Assuming that $h+\eta_0$ is much smaller than $l,$ we can take $\tan\theta\sim\theta,$ and estimate the maximal elevation as
\begin{equation}
\eta_0\simeq\tfrac43a^2h\varepsilon,\end{equation}
which agrees with the result (\ref{elevationkdv}). It follows that $\eta$ can be put in terms of $\eta_0$ as
\begin{equation}
\eta\simeq\eta_0\sech^2a(X+\tfrac43a^2T)
\end{equation}
up to first order. This characterises the form of the free surface traced out by a solitary wave. This was first derived from the tanh-potential (\ref{tanh}) by McCowan \cite{McCowan}, but this connection is more quickly and easily obtained by the simple relation (\ref{elevationkdv}), given a solution to (\ref{ckdv}) or (\ref{pkdv}).
\section{Conclusions}
In this paper, we derived the complex KdV equation for shallow water waves.	We showed that it describes a small perturbation of the complex velocity around steady flow, for the slowly-varying propagation of long waves in a shallow, incompressible, inviscid fluid. An important conclusion of our derivation is that if the complex KdV equation describes a complex velocity for a shallow fluid, then the elevation must be given by a real solution of the KdV equation. In other words, the theory of the real KdV equation is just a particular case of what we have derived here.\\\indent
We reproduced the periodic waves in terms of elliptic functions and the well known result for the soliton solution. However, the complex KdV equation allowed us to calculate not only the surface wave profile, but also the trajectories of particles within the fluid, and show that particles in cnoidal waves move along bean curve-like trajectories. Our results were in close agreement with numerical work based on the real KdV equation \cite{Borluk}, but had the bonus of being a natural and easy consequence of the theoretical framework. This gives even more evidence for the validity of the link between the two approaches established in the present work.\\\indent
Having complete information about the dynamics of all fluid particles, not only at the surface but also throughout the whole body of fluid, may make it possible to solve more complicated problems. For example, this may include the class of problems related to internal waves when the fluid is stratified, or may allow the techniques of complex analysis \cite{Joukowsky} to be applied, such as conformal transformations, which may help to solve problems involving special bottom profiles or underwater barriers and obstacles.
For simplicity, we restricted ourselves to only one type of boundary condition at the bottom, with zero friction. We have no doubt this can be generalised to other conditions that will result in more general solutions. Generalisation to higher-order corrections is also straightforward in principle. The choices here have no limits.\\\indent
We should also mention that the complex KdV equation has a greater variety of solutions  than the real one. This includes regular \cite{Zhang,Yuan,Miller}, blow-up \cite{Charles,Birnir} or complexiton \cite{Chen} solutions. This expands the range of phenomena that the complex KdV equation may describe. On the other hand, we stress that exact solutions, even if they are highly involved, can be found using well known techniques such as the Darboux transformation.\\\indent
Applications of the complex KdV equation are not limited to water waves. This equation can be also used in the seemingly unrelated problem of unidirectional crystal growth \cite{Kerszberg}. Last but not least, the theory of shallow water waves is spreading into the optical domain \cite{Wabnitz} which could be another area of future application.	
\section*{Appendix}
For completeness, we give the results of the calculations for second order perturbations. The second order terms $\beta_2$ in a shallow fluid satisfy the equation
\begin{eqnarray}\nonumber
	\frac{\partial\beta_2}{\partial T}=\beta_1^{\;2}\frac{\partial\beta_1}{\partial Z}+\frac{\beta_1}{18}\frac{\partial^3\beta_1}{\partial Z^3}-\frac{5}{18}\frac{\partial\beta_1}{\partial Z}\frac{\partial^2\beta_1}{\partial Z^2}+\\
	+\frac19\frac{\partial^3\beta_2}{\partial Z^3}-3\frac{\partial(\beta_1\beta_2)}{\partial Z}.
	\label{app}
\end{eqnarray}
where $\beta_1=\beta_1(Z,T)$ is a solution of (\ref{ckdv}).
Having $\beta_1$ in explicit form provides the possibility for higher-order extensions of the present theory.
That is,  working to higher order in perturbation theory will result in equations which include nonlinear and dispersive effects for shallow fluids, accurate to higher powers of the small parameter $\varepsilon=(h/l)^2$. We note here that taking terms which are of higher order in $h/l$ will never result in a theory which is valid for anything other than shallow fluids, which are defined by $h/l$ not being small in the first place.

\end{document}